\documentclass[12pt]{article}
\usepackage{graphicx}
\usepackage{wrapfig}
\RequirePackage{xspace}
\def\lhcb {LHC{\em b\/}\xspace}
\def\lhc {LHC\xspace}

\def\kaon  {\ensuremath{K}\xspace}
\def\epem       {\ensuremath{e^+e^-}\xspace}
\def\g     {\ensuremath{\gamma}\xspace}
\def\B       {\ensuremath{B}\xspace}
\def\btosgam    {\ensuremath{b \to s \g}\xspace}

\def\phigam {\ensuremath{\Bs \to \phi \gamma}\xspace}

\def\kstgam {\ensuremath{\Bd \to \kaon^\ast \gamma}\xspace}

\def\Xs      {\ensuremath{X_s}\xspace}

\def\adelta {\ensuremath {\mathcal{A}^{\Delta}}\xspace}
\def\S {\ensuremath{\mathcal{S}}\xspace}
\def\C {\ensuremath{\mathcal{C}}\xspace}
\def\Bd      {\ensuremath{B_d}\xspace}
\def\Bs      {\ensuremath{B_s}\xspace}
\def\invfb   {\ensuremath{\mbox{\,fb}^{-1}}\xspace}

\newcommand\pubnumber{LHCb-PROC-2011-003}
\newcommand\pubdate{\today}
\def\napoli{Department of Physics, Blackett Lab,\\
Imperial College London, SW7 2AZ, United Kingdom}
\def\Title#1{\begin{center} {\Large #1 } \end{center}}
\def\Author#1{\begin{center}{ \sc #1} \end{center}}
\def\Address#1{\begin{center}{ \it #1} \end{center}}

\newcommand\pubblock{\rightline{\begin{tabular}{l} \pubnumber\\
         \pubdate  \end{tabular}}}
\newenvironment{Abstract}{\begin{quotation}  }{\end{quotation}}
\newenvironment{Presented}{\begin{quotation} \begin{center} 
             PRESENTED AT\end{center}\bigskip 
      \begin{center}\begin{large}}{\end{large}\end{center} \end{quotation}}
\def\Acknowledgements{\bigskip  \bigskip \begin{center} \begin{large}
             \bf ACKNOWLEDGEMENTS \end{large}\end{center}}
\def\beq{\begin{equation}}
\def\eeq#1{\label{#1}\end{equation}}
\def\eeqn{\end{equation}}

\def\beqa{\begin{eqnarray}}
\def\eeqa#1{\label{#1}\end{eqnarray}}
\def\eeqan{\end{eqnarray}}

\let\bar=\overbar

\def\Dslash{\not{\hbox{\kern-4pt $D$}}}
\def\dslash{\not{\hbox{\kern-2pt $\del$}}}

\def\msb{{\bar{\ssstyle M \kern -1pt S}}}

\begin{document}
\begin{titlepage}
\pubblock

\vfill
\Title{$B \to X_{s} \gamma$ and $B \to X_{s} l^{+} l^{-}$ decays at LHCb}
\vfill
\Author{Fatima Soomro \footnote{On behalf of the \lhcb collaboration}}
\Address{\napoli}
\vfill
\begin{Abstract}
\lhcb is one of the four major experiments at the Large Hadron Collider (\lhc) at CERN. It is custom built to look for CP violation and New Physics in rare decays of heavy flavour hadrons, like the B and D systems. Rare decays that occur via loop diagrams provide a way to probe New Physics at energy scales much higher than can be probed by direct production in experiments. In this article, the \lhcb prospects for such measurements with exclusive decays of the type $B \to \Xs \gamma$ and $B \to X_{s} l^{+} l^{-}$ are presented.
\end{Abstract}
\vfill
\begin{Presented}
6th International Workshop on the CKM Unitarity Triangle\\
Warwick, United Kingdom, September 6-10, 2010
\end{Presented}
\vfill
\end{titlepage}
\setcounter{footnote}{0}

\section{Introduction}
The transition of a \emph{b} quark into an \emph{s} quark is a flavour changing neutral current, and is forbidden at tree level in the Standard Model (SM). It can only proceed via loop diagrams, an example of which is shown in Fig. 1. Decays involving such transitions are excellent probes of beyond SM physics.\\

\begin{figure}[h!]
\centering
\includegraphics[width=0.99\textwidth, trim=0 130mm 0 0, clip]{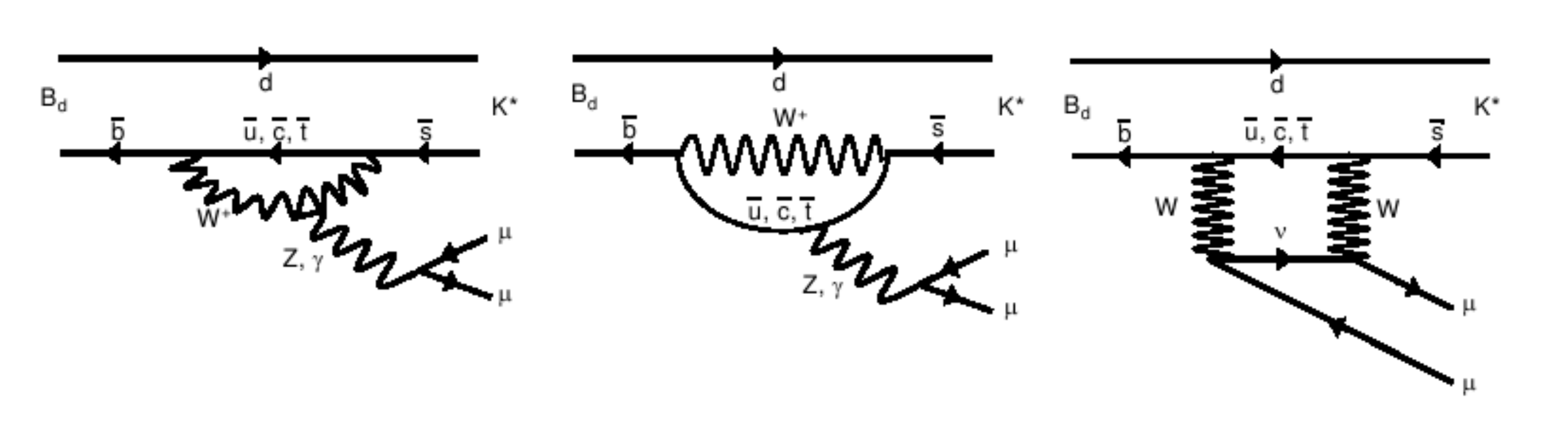}
\label{fig:1}
\caption{Feynman diagrams corresponding to $\Bd \to K^{*} \gamma$ and $\Bd \to K^{*} \mu^{+} \mu^{-}$. The diagram to the right is exclusive to $\Bd \to K^{*} \mu^{+} \mu^{-}$ while the one at the left and centre can represent a $\Bd \to K^{*} \gamma$ decay if the final state electroweak boson is an unconverted photon.}
\end{figure}
The SM prediction of the inclusive rate $B(B \to X_{s} \gamma) = (3.15 \pm 0.23) \times 10^{-4} $ and the experimental value $(3.56\pm0.26)\times10^{-4} $~\cite{PDG, BelleIncMeas} are in good agreement, which can be used to put constraints on various NP models. An example of this is shown in Fig.2.\\ 

\vspace{-5mm}
\begin{figure}[h]
\begin{minipage}[l]{0.45\textwidth}
 \begin{center}
    \includegraphics[width=1.0\textwidth, trim = 50mm 30mm 40mm 28mm, clip]{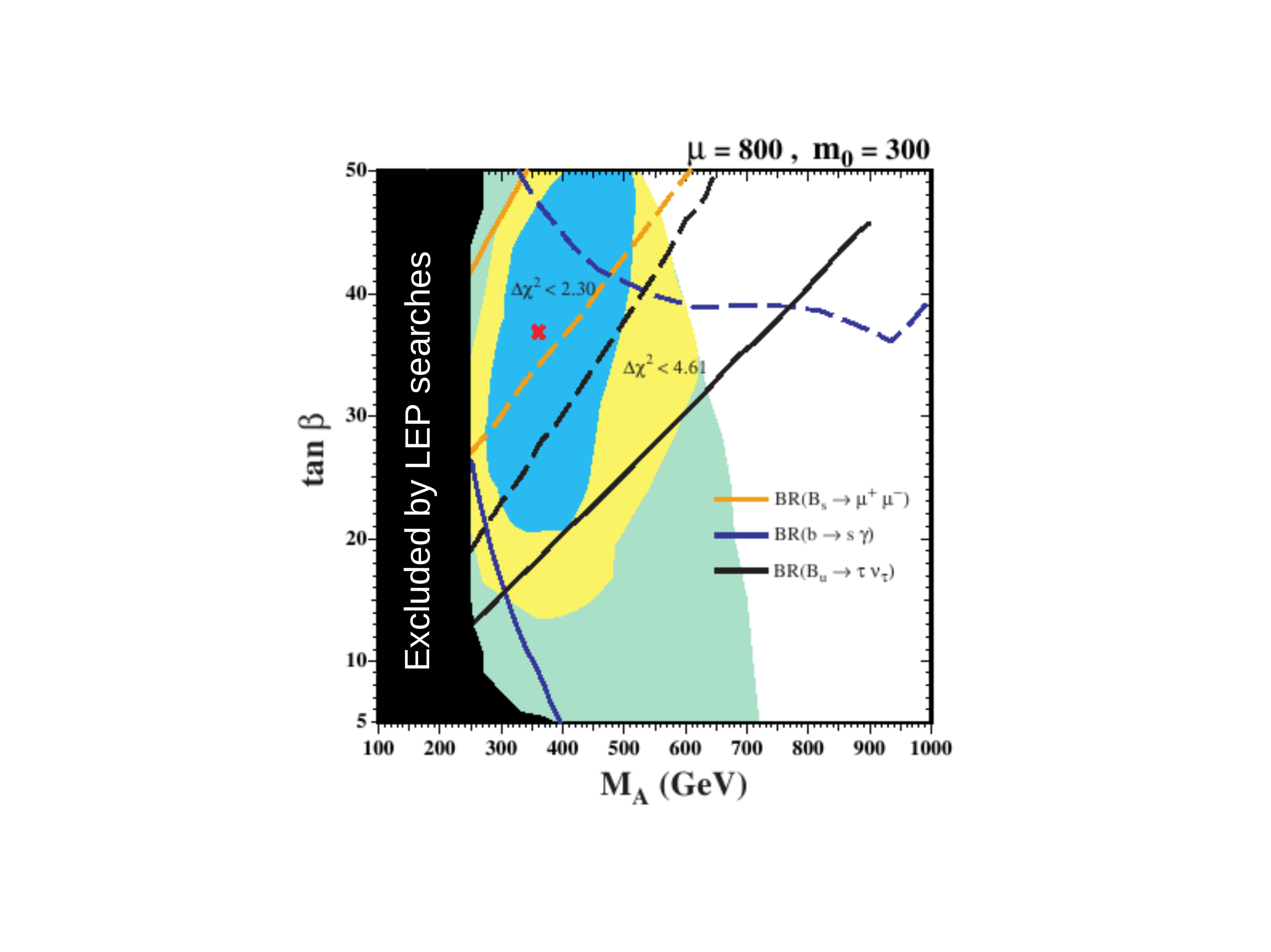}
  \end{center}
  %\caption{Exclusion regions in the ($M_A, tan \beta$) plane for a certain MSSM scenario (taken from ~\cite{figureRef}) }
\end{minipage}
\begin{minipage}[r]{0.55\textwidth}
{\small{Figure 2: Exclusion regions in the ($\mathrm{M}_{\mathrm{A}}, \mathrm{tan \beta}$) plane for the Non Universal Higgs Model (NUHM) realisation of the Minimal Supersymmetric Standard Model (MSSM), taken from~\cite{figureRef}. The coloured contours show the 1 and 2 $\sigma$ allowed regions for the best fit value, indicated by the red cross. \\The various lines show the sensitivites of the observables BR($\Bs \to \mu \mu$), BR(\btosgam) and BR($\B_{u} \to \tau \nu$)}}%The blue solid (dashed) line shown the exclusion area for BR(\btosgam) = 4(3)$\times 10^{-4}$.}}
\end{minipage}
\end{figure}
\addtocounter{figure}{1}
Similar agreement has been established between SM predictions and experimental measurements for the exclusive $B \to X_{s} \gamma$ and the inclusive and exclusive $B \to X_{s} l^{+} l^{-}$ branching ratios~\cite{ali, HFAG}. However, NP can show up in observables other than decay rates, such as polarizations of the final state particles and asymmetries in their angular distributions. The polarization of the photon in $B \to \Xs \gamma$ decays and the angular distributions in $B \to X_{s} l^{+} l^{-}$ are discussed in this article, in the context of the \lhcb detector.
\section{Photon polarization}
The photon emitted in a \btosgam transition is predominantly left handed due to the helicity structure of the weak interaction. There is a small right handed component due to the finite \emph{s} quark mass, and the ratio of the right handed and left handed components, $A_R/A_L$, is proportional to the ratio of the quark  masses $m_s/m_b$. In the SM, this ratio has been predicted to be about 0.4\% while it can be enhanced by up to 10\% in some NP models~\cite{soni}.\\
The polarization of the photon in \btosgam decays can be measured in several channels~\cite{Methods}, and at \lhcb, analyses with \phigam and $B \to K^{*} \epem$ have been developed to make this measurement \footnote{Here and later in this report, the $K^{*}$ from the \B decay is reconstructed as $K^{*} \to K \pi$ unless noted explicitly.}. In such decays, there is interference in mixing and decay. For example, a $\bar{\B_s}$($\B_s$) decays predominantly to a left (right) handed photon, but can also decay to a right (left) handed one, although the latter final state is suppressed for a $\bar{\B_s}$($\B_s$).
\subsection{\phigam at \lhcb}
The photon polarization measurement can be made by looking at the time dependent decay rate of $B \to f_{CP}\gamma$ decays, which can be expressed as
\begin{eqnarray}
    \Gamma_{ \mathrm{B(\bar{B})} \to {\mathit{f}}^{\mathcal{CP}}\gamma}
    \left( \mathrm{t}\right)=
    \left|\mathrm{A}\right|^2\mathrm{e}^{-\Gamma \mathrm{t}}
    \bigl(\cosh \frac{\Delta \Gamma \mathrm{t}}{2}\;-\;
    \mathcal{A}^{\Delta} \sinh \frac{\Delta \Gamma \mathrm{t}} {2}\; \pm \;
      \mathcal{C}\cos\Delta \mathrm{m} \mathrm{t}\; \mp \;
      \mathcal{S}\sin\Delta \mathrm{m} \mathrm{t}\bigr)
      \label{decrate} 
  \end{eqnarray}
and the parameters \S and \adelta can be expressed as 
\begin{equation}
\mathcal{S} \approx \sin 2\psi\sin\varphi_{\mathrm{s}}, \quad \mathcal{A}^{\Delta}  \approx \sin 2\psi\cos\varphi_{\mathrm{s}}\quad \mathrm{and} \quad \tan\psi 
\equiv \frac{A_R}{A_L} = 
\frac{ \mathcal{A}\left(\bar{\mathrm{B}}_{\mathrm{s}}\rightarrow \mathit{f}^{\mathcal{CP}}\gamma_{\mathrm{R}}\right)}
{\mathcal{A}\left(\bar{\mathrm{B}}_{\mathrm{s}}\rightarrow \mathit{f}^{\mathcal{CP}}\gamma_{\mathrm{L}}\right)} 
\end{equation}
which contain $\psi$, the parameter sensitive to the ``wrong'' photon polarization fraction. The parameter \S has been measured at the \emph{B} factories, from the time dependent decay rate of $\B_d \to K^* (K_s \pi^0)\gamma$. The current average is $\S = -0.19 \pm 0.23$~\cite{HFAG}.\\
At \lhcb, \phigam will be used to measure the photon polarization parameters \S and \adelta. The \Bs system has sensitivity to \adelta because of the large $\Delta \Gamma_s$~\cite{widthDiff}. The resolution on \adelta is inversely proportional to the value of $\Delta \Gamma_s$.\\
The measurements of \adelta and \S are very complementary as these parameters appear with the \emph{cosine} and \emph{sine} of $\phi_s$ respectively, where $\phi_s$ is the \Bs mixing phase. If $\phi_s$ is small, as the SM predicts, the measurement of \adelta will have better sensitivity on $\sin 2\psi$. Conversely, sensitivity on $\sin 2\psi$ will come from a measurement of \S if $\phi_s$ is enhanced by NP.\\
The measurement of \adelta does not require flavour tagging, because the terms containing \S and \C will cancel in the decay rates of \Bs and $\bar{\Bs}$ mesons.
With a data set corresponding to 2\invfb at \lhcb, about 11000 reconstructed and selected \phigam events are expected, with a background to signal ratio (\emph{B/S}) of $<$ 0.55~\cite{roadmap}. The sensitivity on $\sin 2\psi$ is expected to be $\sim$ 0.2 for the tagged analysis and $\sim$ 0.22 for the untagged one. The latter analysis is briefly described below.

%\hspace{-10mm}\textbf{Untagged analysis of \phigam for \adelta}\newline
\subsubsection{Measurement of \adelta}
For small values of \adelta, the decay rate (Eq. \ref{decrate}) becomes $\Gamma_{ \mathrm{B} \to {\mathit{f}}^{\mathcal{CP}}\gamma}\left( \mathrm{t}\right)\approx \left|\mathrm{A}\right|^2\mathrm{e}^{-\Gamma_{\phigam} \mathrm{t}}$, where $\Gamma_{\phigam} = \Gamma+\frac{\adelta\Delta\Gamma}{2}$. Therefore, the measurement of \adelta is actually a measurement of the difference in the \Bs proper time as measured in \phigam and in some other channel. In order to make a precision measurement, any bias in the proper time reconstruction needs to be understood to a few precent of the \Bs proper time.\\
\vspace{-5mm}
\begin{wrapfigure}{r}{0.5\textwidth}
\vspace{-10mm}
 \begin{center}
    \includegraphics[width=0.5\textwidth, trim = 0 0 0 12mm, clip]{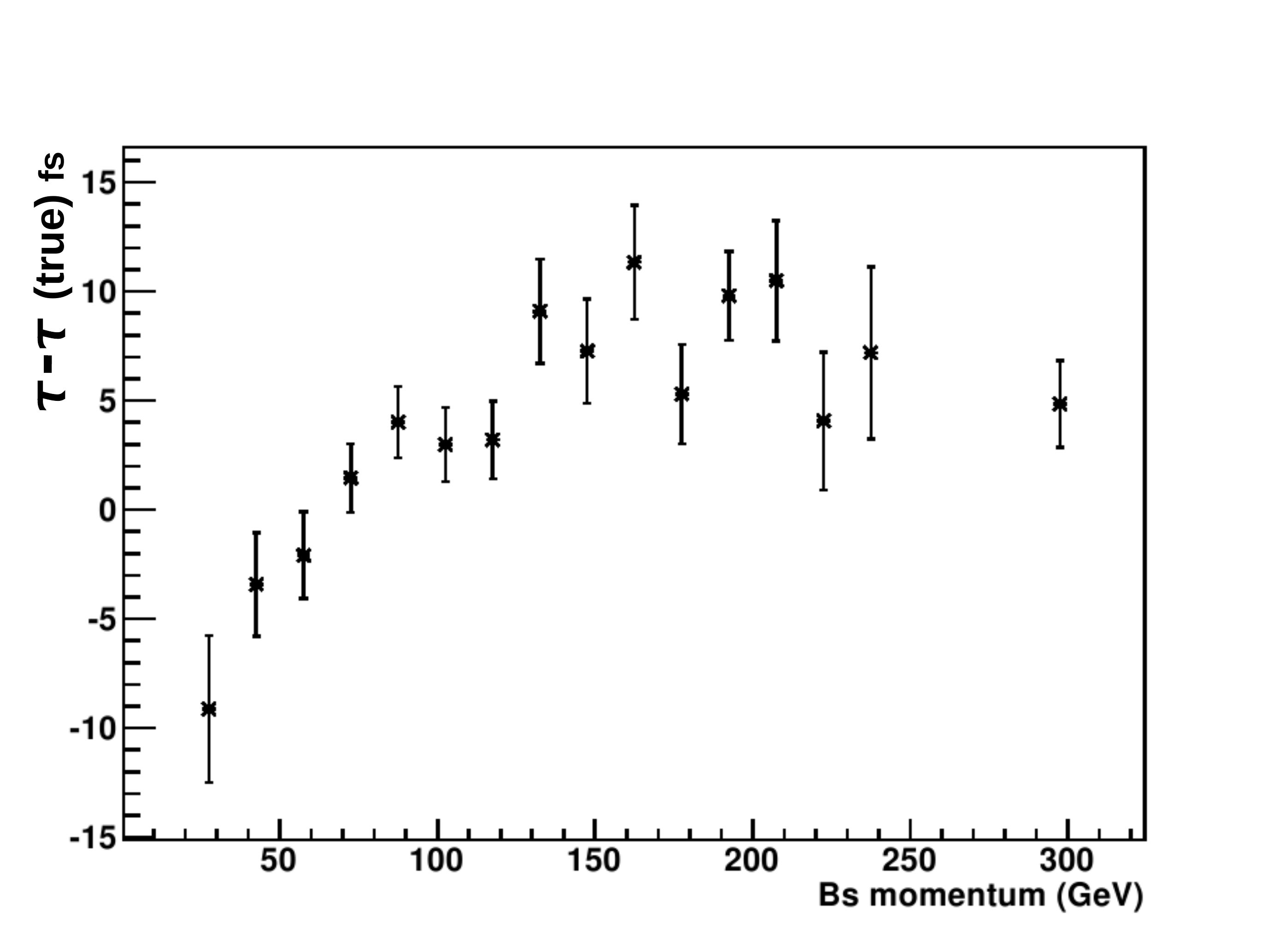}
  \end{center}
\vspace{-8mm}
\caption{{\small{Bias in the proper time reconstruction, in $\mathrm{fs}$, as a function of the \Bs momentum.}}}
\label{fig:3}
\end{wrapfigure}
From a MC signal sample, the bias in the \Bs proper time reconstruction is plotted as a function of the \Bs momentum, in Fig.~\ref{fig:3}. This bias is due to a bias in the \Bs momentum reconstruction, the error on which is dominated by the photon momentum reconstruction.\\
From toy MC studies, it is estimated that a bias of this magnitude can introduce a considerable shift in the measured value of \adelta, especially if $\Delta\Gamma_s$ is measured to be larger, enhancing the \adelta sensitivity. This makes calorimeter calibration a crucial requirement of the analysis. The decay \kstgam, with an expected yield six times greater than that of \phigam, is a good control channel for making this calibration.\\
Another important ingredient of this analysis is the efficiency to reconstruct a given proper time, or the proper time acceptance function. From MC studies, it has been established that this acceptance also needs to be determined to a few percent level, at long \Bs proper times. The decay $\Bs \to \phi J/\psi$ can be used as a control channel to extract the \phigam proper time acceptance function.

\subsection{$\Bd \to K^{*} \epem$ at \lhcb}
\label{sec:kstee}
\begin{figure}[h]
\begin{minipage}[l]{0.5\textwidth}
  \begin{center}
    \includegraphics[width=0.95\textwidth, trim=80mm 50mm 60mm 50mm, clip]{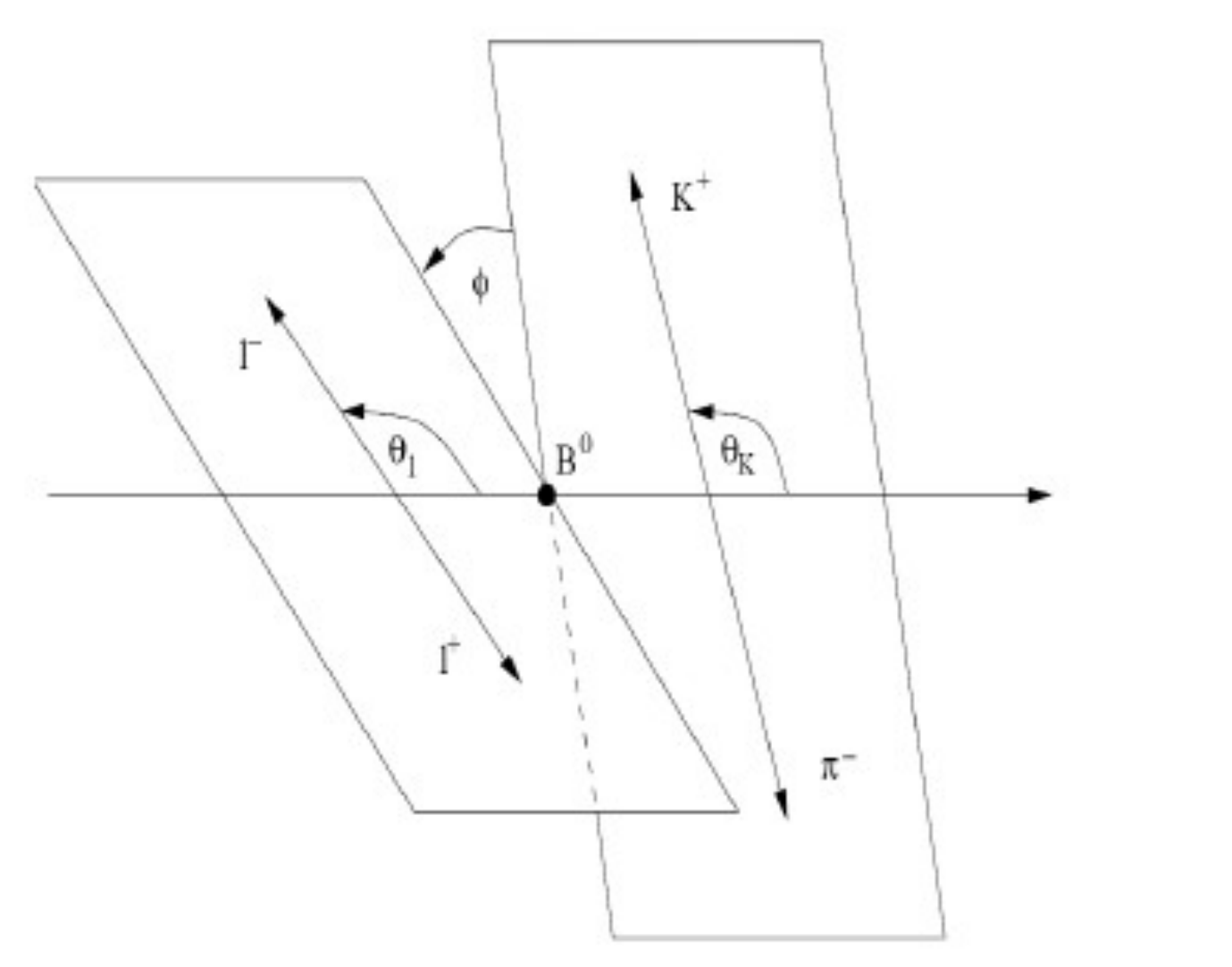}
  \end{center}
  \vspace{-6mm}
  \caption{{\small{Definition of the various angles in $\Bd \to K^{*} \epem$}}}
  \label{fig:5}
\end{minipage}
\hspace{2mm}
\begin{minipage}[r]{0.5\textwidth}
  \begin{center}
    \includegraphics[width=0.9\textwidth, trim=0mm 50mm 100mm 0mm, clip]{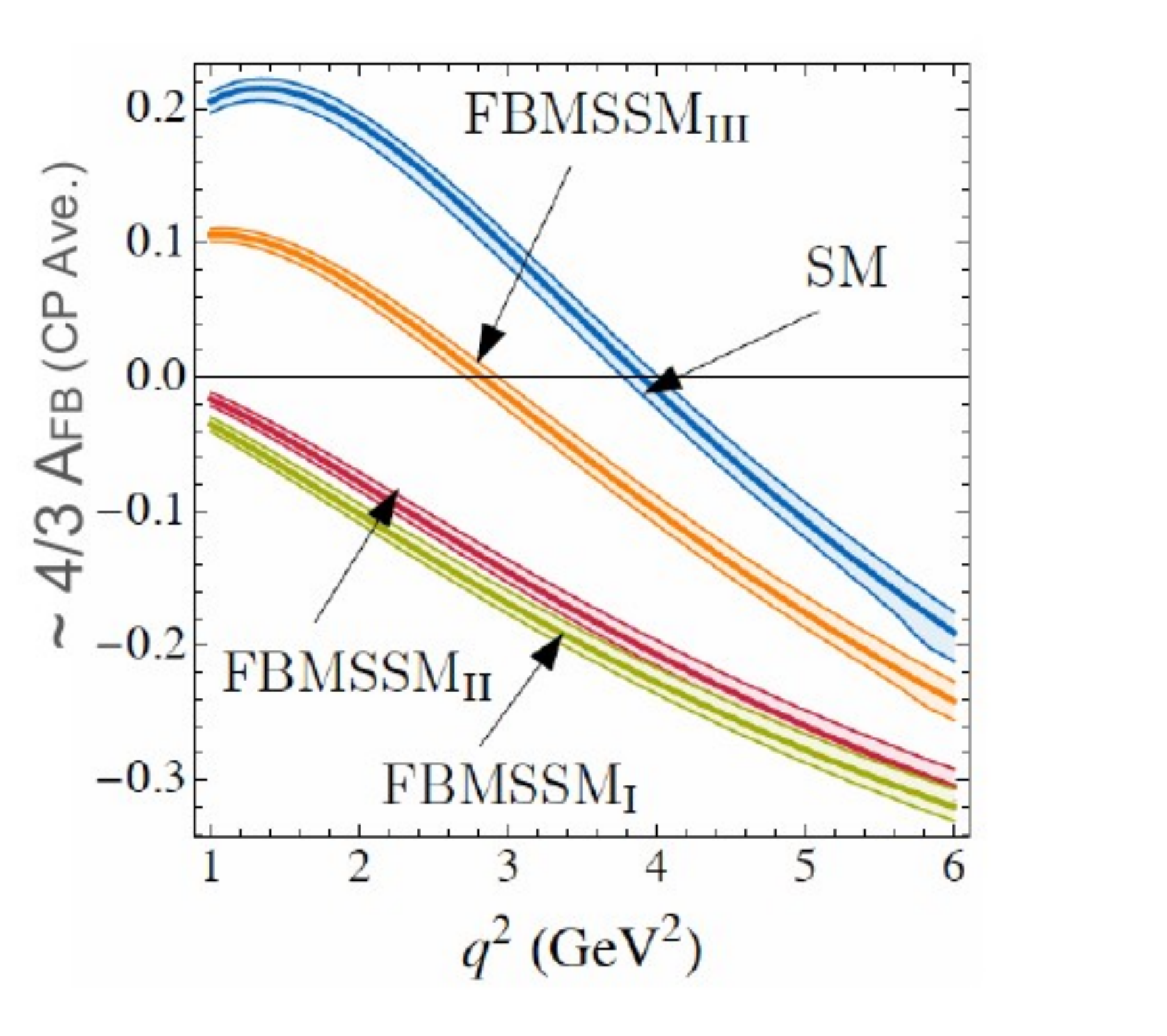}
  \end{center}
  \vspace{-10mm}
  \caption{{\small{$A_{FB}$ as a function of $q^2$, in the SM and some NP models}}}
  \label{fig:6}
\end{minipage}
\end{figure}

In $\Bd \to K^{*} \epem$ decay, the photon polarization observable $A_T ^{(2)} = -2 \frac{A_R}{A_L}$ can be extracted from a fit to the angular distributions of the decay~\cite{Matias}. At very low di electron invariant mass squared, $q^2$, the $e^+ e^-$ production is dominated by a virtual photon, and the measurement in $\Bd \to K^{*} \epem$ is equivalent to the polarization measurement in \phigam. The two key angles in $\Bd \to K^{*} \epem$ are the angle ($\phi$) between the decay planes of the electron pair and the $K$ and $\pi$ from the $K^*$, and the angle ($\theta_K$) between the $K$ and the \Bd flight direction, as shown in Fig.~\ref{fig:5}.\\
From a data set of 2\invfb, around 200 reconstructed and selected signal events are expected with a \emph{B/S} $\sim$ 1. The statistical sensitivity on the ratio of left and right amplitudes, $A_R/A_L$, is estimated to be 0.12. From MC studies, it has been established that the analysis is robust against angular biases, and systematic uncertainties are expected to be small~\cite{Kstee}.
\section{Angular observables in $\Bd \to K^* \mu^{+} \mu^{-}$ at \lhcb}
The angular definitions of Fig.~\ref{fig:5} hold for $\Bd \to K^*\mu^{+} \mu^{-}$ as well and its phenomenology is also very similar to $\Bd \to K^{*} \epem$. However, the higher expected yield for the dimuon final state allows access to a variety of angular observables~\cite{Ulrik}. An interesting observable is the forward backward asymmtery ($A_{FB}$), which is defined as the difference between the number of positive and negative leptons going in the same direction as the \emph{s} quark, in the dilepton rest frame. Different NP scenarios predict different shapes of $A_{FB}$ as a function of $q^2$, as shown in Fig.~\ref{fig:6}~\cite{NPinAFB}.\\
%Inclusive $B \to X_s l^{+} l^{-}$ decays are hard to measure experimentally, while for exclusive decays, the theoretical predictions have large hadronic uncertainties.\\
%One observable for which the hadronic uncertainties cancel is the \emph{zero crossing point}, which is the point along $q^2$, where $A_{FB}$ is zero. It has clean theoretical predictions from the SM and also from NP models. 
The measurements of $A_{FB}$ by the \emph{B} factories and CDF \cite{afbMeasurements} are limited by statistics. Nevertheless, there is an interesting hint of a deviation from the SM. Assuming that the Belle central value in the range $1 < q^2 < 6$ holds, \lhcb can exclude the SM at 5$\sigma$ level with a data sample corresponding to 2\invfb.

\subsection{\lhcb detector performance} 
\begin{figure}[h]
\begin{minipage}[l]{0.5\textwidth}
  \begin{center}
    \includegraphics[width=0.94\textwidth, trim= 0 0 8mm 5mm , clip]{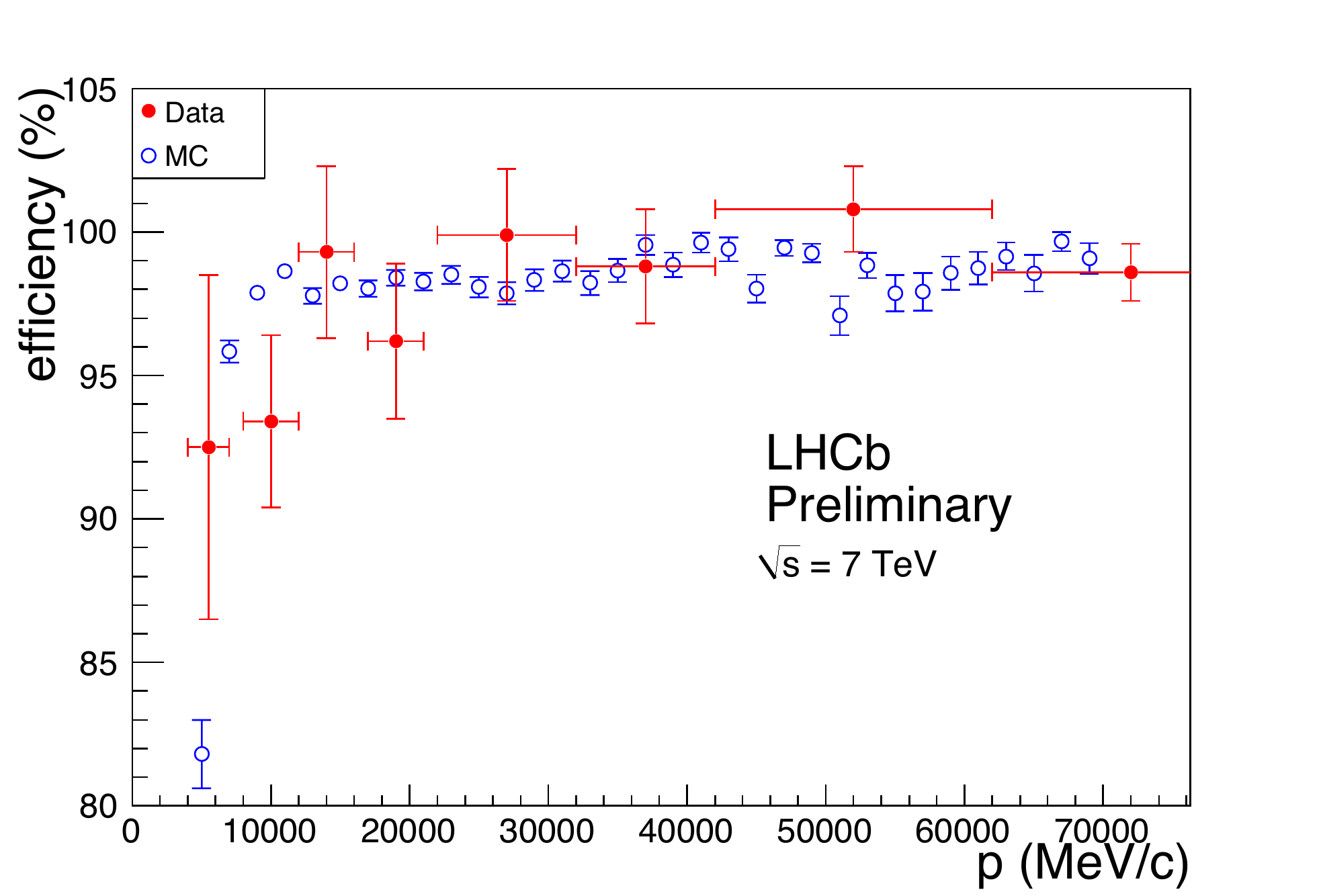}
  \end{center}
  \vspace{-6mm}
  \caption{{\small{The efficiency to positively identify a muon, as a function of the muon momentum, from data and Monte carlo}}}
  \label{fig:7}
\end{minipage}
\hspace{2mm}
\begin{minipage}[r]{0.5\textwidth}
  \vspace{3mm}
  \begin{center}
    \includegraphics[width=0.92\textwidth, trim= 0 0 0 3mm, clip]{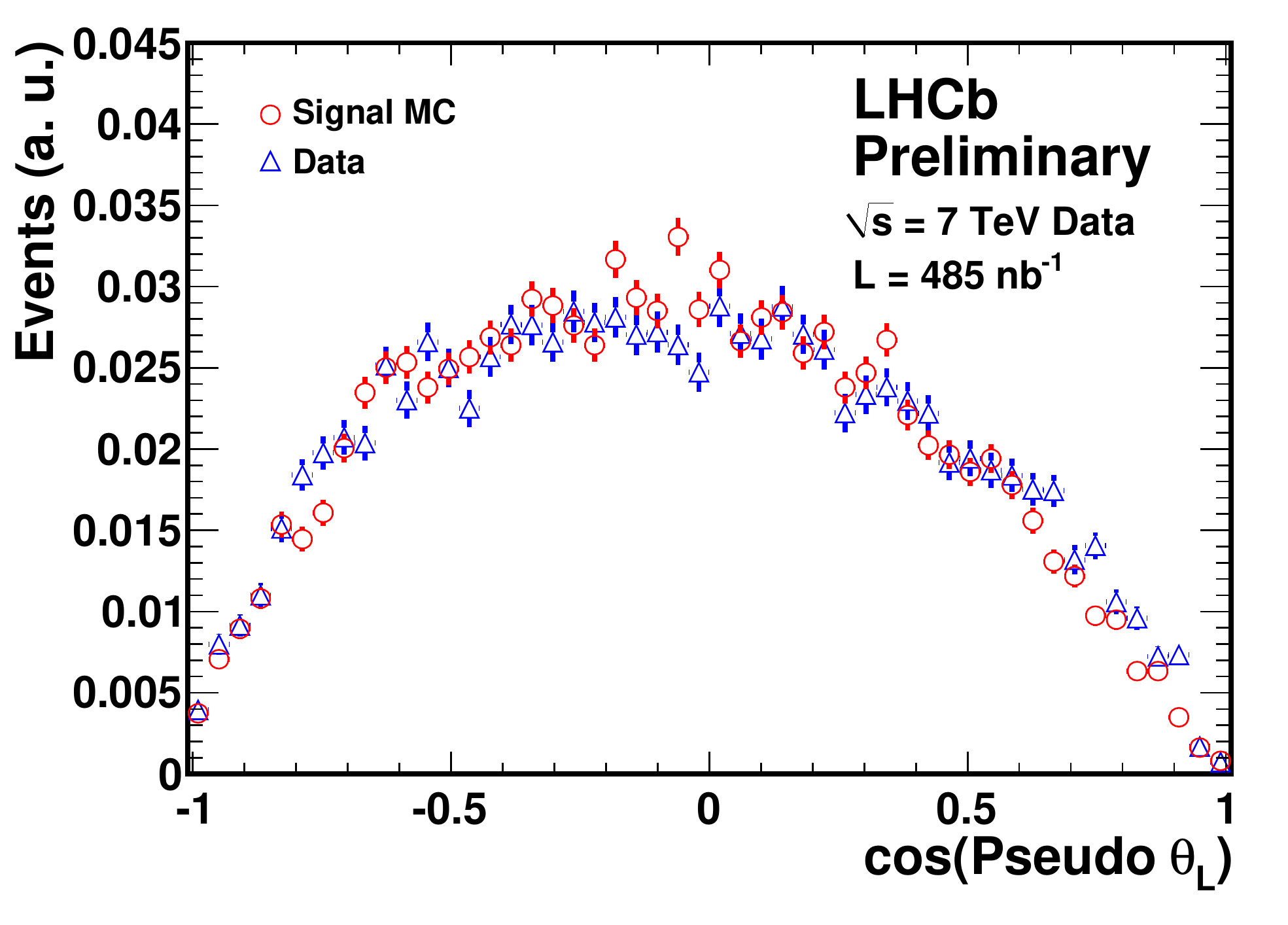}
  \end{center}
  \vspace{-9mm}
  \caption{{\small{Angular distribution of $D \to K \pi \pi \pi$ from data, compared to the signal Monte carlo prediction}}}
  \label{fig:8}
\end{minipage}
\end{figure}

The performance of the subdetectors which are relevant to the analyses of rare decays has been validated with the data collected in 2010. A preliminary example for the muon system is shown in Fig.~\ref{fig:7}. This shows the efficiency to positively identify a muon, as a function of its momentum. The efficiency in data has been measured by using the tag and probe method on muons from $J/\psi$ decays and agrees very well with the MC prediction.

\vspace{-4mm}
The measurement of angular variables in $\Bd \to K^* \mu^{+} \mu^{-}$ requires good control of angular biases due to detector acceptance, and selection criteria. A four body final state such as $D \to K \pi \pi \pi$ can be used as a proxy to test the angular biases. Fig.~\ref{fig:8} shows the angular distribution of $D \to K \pi \pi \pi$ from data and MC, which demonstrate excellent agreement between simulation and data.% Similarly, the yield of $\Bd \to K^* J/\psi$ is compatible with MC estimates of selection efficiency.
\section{Conclusions and outlook}
This article presents an overview of the \lhcb measurements of the photon polarization in \phigam and $\Bd \to K^* e^+ e^-$ decays, and of the angular observables in $B \to K^* \mu^+ \mu^-$ decays.\\
For the photon polarization measurement, at least 2\invfb of data are required in order to improve on the current sensitivity on the \adelta parameter. With $B \to K^* \mu^+ \mu^-$ decay, \lhcb will be able to make the world's best measurement of $A_{FB}$. Assuming that the Belle central value of $A_{FB}$ in the range $1 < q^2 < 6$ holds, \lhcb can establish a discrepancy wrt the SM prediction at 5$\sigma$ level, with 2\invfb of data.\\
%\lhcb can make a competitive measurement of $A_{FB}$ in , with only 100\invpb. With 2\invfb, the \lhcb measurement of $A_{FB}$ has the potential to make a 5$\sigma$ exclusion of the SM. By the end of 2010, \lhcb accumulated 40\invpb of integrated luminosity and of the order of 1\invfb are foreseen to be collected during 2011, making a world class measurement of $A_{FB}$ feasible at \lhcb with next year's data. Other angular observables have also been explored theoretically, and those analyses will need more data and understanding of the angular biases introduced by the detector acceptance and selection cuts.\\
With the data collected in 2010, the performance of various subdetectors has been determined and shown to agree well with simulation. The selection criteria and efficiency and angular bias predictions are also being validated using control channels and have shown promising results. The \lhcb collaboration is looking forward to the start of data taking in 2011, during which the detector is foreseen to accumulate $\sim$1\invfb of data, which will be followed by exciting results from these analyses.

\end{document}